\documentclass{article}
\usepackage{ijcai16}

\usepackage{times}
\usepackage{graphicx}
\usepackage{subfigure}
\usepackage{url}

\usepackage{algorithmic,tikz,rotating}
\usetikzlibrary{arrows,shapes,snakes}

\usepackage{comment} 

\title{A Search/Crawl Framework for Automatically Acquiring Scientific Documents}
\author{Sujatha Das Gollapalli$^1$, Krutarth Patel$^2$, and Cornelia Caragea$^2$ \\ 
$^1$Institute for Infocomm Research, A*STAR, Singapore \\
$^2$Department of Computer Science and Engineering, University of North Texas, Denton, TX  \\
{\tt gollapallis@i2r.a-star.edu.sg,KrutarthPatel@my.unt.edu,ccaragea@unt.edu}}

\begin{document}

\maketitle
\begin{abstract}
Despite the advancements in search engine features, ranking methods, technologies, 
and the availability of programmable APIs,
current-day open-access digital libraries still rely on crawl-based approaches
for acquiring their underlying document collections. In this paper,
we propose a novel search-driven framework for
acquiring documents for scientific portals. % such as CiteSeer$^{\tt x}$.
Within our framework, publicly-available research paper titles and author names 
are used as queries to a Web search engine. Next, research papers and sources of 
research papers are identified from the search results using accurate
classification modules. %We describe our framework and
%features used in designing our classifiers. 
Our experiments highlight
not only the performance of our individual classifiers but also 
the effectiveness of our overall Search/Crawl framework. Indeed, we
were able to 
obtain approximately $0.665$ million research documents 
through our fully-automated framework using
about $0.076$ million queries. These prolific results 
position Web search as an effective alternative to crawl methods
for acquiring both the actual documents and
seed URLs for future crawls.
\end{abstract}
%\vspace{-12pt}
%comment-cc: could you please check the template. this format here is different (smaller font and larger margins) compared with the regular AAAI/IJCAI template.
\section{Introduction}
\label{sec:intro}
Scientific portals such as PubMed, Google Scholar, Microsoft Academic Search, 
CiteSeer$^{\tt x}$, and ArnetMiner provide access to scholarly publications and 
comprise indispensable resources for researchers who search for literature 
on specific subject topics. In addition, data mining applications such as 
citation recommendation~\cite{wsdm11he},  expert search~\cite{ijcai07balog},
topic trend detection~\cite{kdd06wang,cikm09he}, and author influence modeling~\cite{ijcai11kataria}
involve web-scale analysis of up-to-date research collections. While
academics and researchers\footnote{\scriptsize In this paper, we use 
the terms ``researchers/authors/scholars" and ``research documents/papers/publications" interchangeably. 
We also use (academic) homepages to refer to professional homepages maintained by scholars and {``Scholarly/Academic Web" to 
refer to sections of the Web (for example, university websites and research centers) that cater to scholarly pursuits.}} 
 continue to produce large numbers of scholarly documents worldwide, acquisition of 
research document collections becomes a challenging task for digital libraries.

In contrast with commercial portals 
(such as the ACM digital library)
that rely on clean and structured publishing sources for their collections, open-access, autonomous systems such as CiteSeer$^{\tt x}$ and ArnetMiner 
acquire and index freely-available research articles on the Web~\cite{infoscale06li,kdd08tang}. 
%comment-cc: I am skeptical to use this since once Lee said open access journals / conf. don't like when they are crawled. Let's keep it only for homepages and seed URLs.
%Open-access journal and conference websites, 
Researchers' homepages and paper repository URLs are crawled and processed 
periodically for maintaining the research collections in these portals. Needless to say, 
these repositories are incomplete since the crawl seed lists cannot 
be comprehensive in face of the ever changing Scholarly Web. Not only do new authors and publication venues
emerge, but also existing researchers may stop publishing or change universities resulting in
outdated seed URLs. \textit{Given this challenge, how can we automatically augment crawl seed lists
for a scientific digital library?}

Web search has been a constant topic of investigation for
IR, ML, and AI research groups since several years. Current 
Web search engines feature state-of-the-art
technologies, ranking algorithms, syntax, personalization and localization features along
with efficient infrastructure and programmable APIs making them invaluable tools
to access and process the otherwise intractable Web. Despite these attractive advancements, to the best of 
our knowledge, search-driven methods are yet to be investigated as alternatives
to crawl-based approaches for acquiring documents in digital libraries. In this paper, 
we address 
this gap in the context of open-access, scientific digital libraries. We propose
a novel Search/Crawl framework, describe its components and present experiments 
showcasing its potential in acquiring research documents.

To motivate our framework, we recall how a
Web user typically searches for research papers or authors~\cite{nips02richardson,cikm08serdyukov}. 
As with regular document search, a user typically issues Web search queries 
comprising of representative keywords or paper titles
for finding publications on a topic. Similarly, if the author is known, 
a ``navigational query"~\cite{sigir02broder} may be employed to locate the homepage
where the paper is likely to be hosted. Indeed, according to
previous studies, researchers provide access to their papers (when possible) to improve their visibility and 
citation counts making researcher homepages {a likely hub for locating
research papers}~\cite{nature01lawrence,tweb15gollapalli}.
\begin{figure*}[!htp]
\centering
\hspace*{-0.65cm}
\includegraphics[scale=0.35]{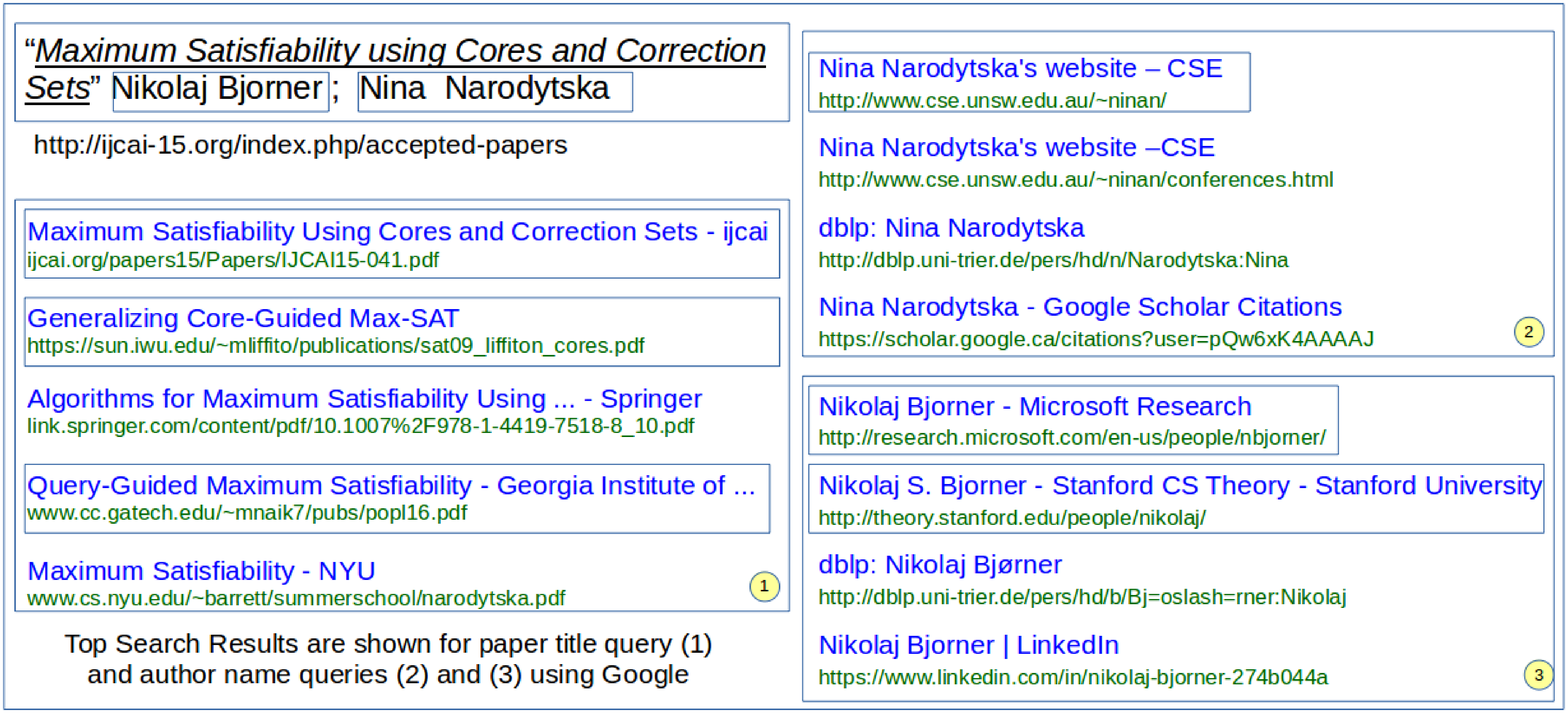}
%\vspace{-10pt}
\caption{\small An anecdotal example for illustration (searches performed on Jan 26, 2016).}
%\vspace{-10pt}
\label{fig:anecdotalsearch}
\end{figure*}

Given previous knowledge 
in academic browsing, scholars are often able to accurately locate the correct research 
papers or academic homepages from the Web search results using hints
from the titles, search summaries (or snippets) and the URL strings. To illustrate this process, Figure~\ref{fig:anecdotalsearch} shows an anecdotal example of a 
search using Google for the title and authors of a paper published at IJCAI last year, ``Maximum Satisfiability using Cores and Correction Sets'' by Nikolaj Bjorner and Nina Narodytska. For the top-$5$ results shown for the paper title query (set 1), 
four of the five results are research papers on the topic. The document
at the Springer link is not available for free whereas the last document corresponds to
course slides.
For the homepage URLs identified from author name search results (from sets 2 and 3), namely: \\
\begin{scriptsize}
\texttt{\textbf{http://www.cse.unsw.edu.au/~ninan/}} \\
\texttt{\textbf{http://research.microsoft.com/en-us/people/nbjorner/}} \\
\texttt{\textbf{http://theory.stanford.edu/people/nikolaj/}} 
\end{scriptsize}
\\
we found $55$ documents, $46$ of which correspond to research publications.
%%URL #PDF/PS/PS.* #papers #uniqueTitles
%%http://www.cse.unsw.edu.au/~ninan/     1    0    0    
%%http://research.microsoft.com/en-us/people/nbjorner/     39    34    34   
%%http://theory.stanford.edu/people/nikolaj/     16    12    11 
This anecdotal search example highlights the immense potential of Web search 
for retrieving research papers and seed URLs that can be crawled for research papers.

\begin{figure}[!htbp]
\centering
\includegraphics[height=2.5in, width=3in]{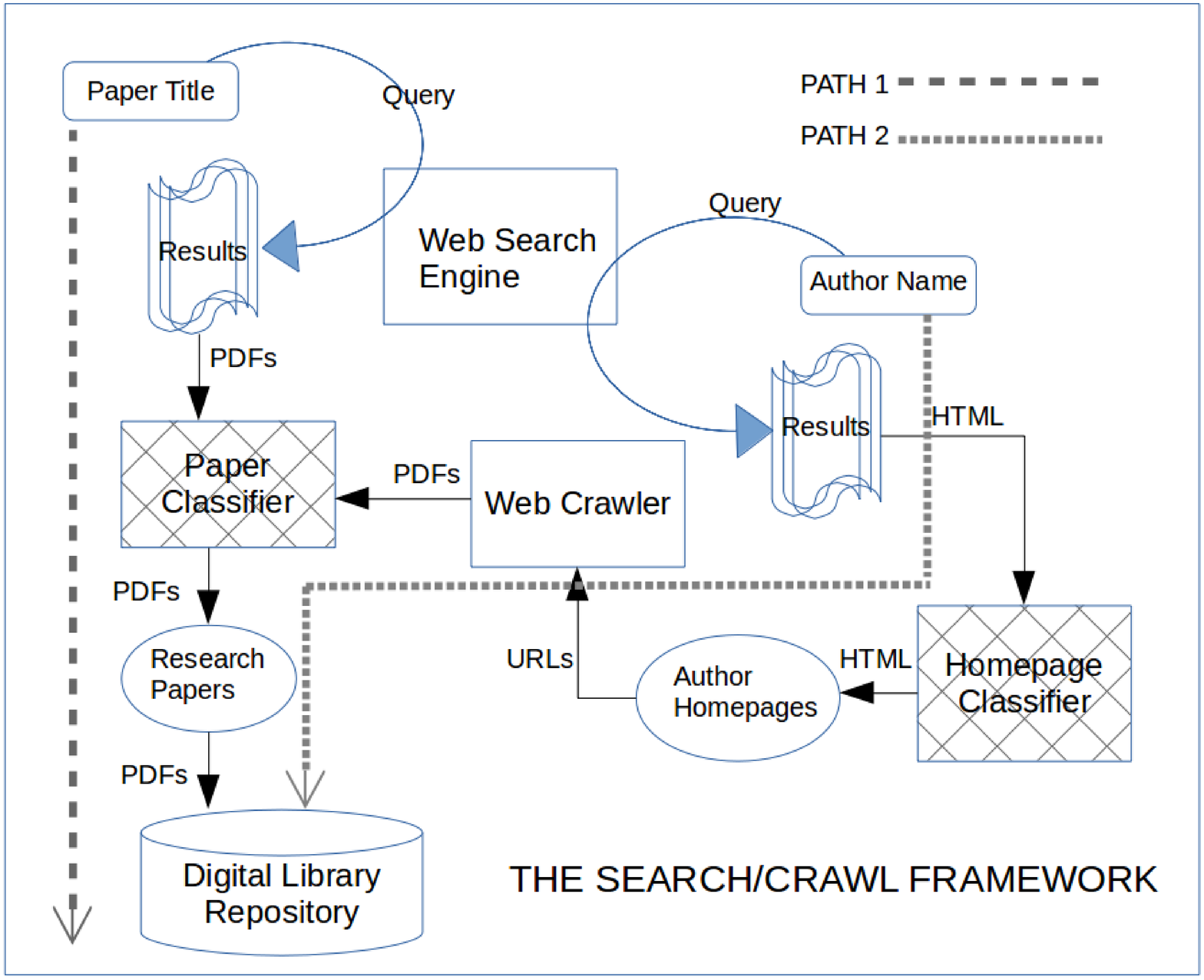}
%\vspace{-10pt}
\caption{\small Schematic Diagram of our Search/Crawl framework.}
\label{fig:schematic}
\end{figure}

Our Search/Crawl framework mimics precisely the above search and scrutinize approach
adopted by Scholarly Web users. Freely-available information from the Web for specific subject disciplines\footnote{
\scriptsize For example, from bibliographic listsings such as DBLP.} % and conference/journal websites}
is used to frame title and author name queries in our framework. The two control flow paths
for obtaining research papers are highlighted in 
Figure~\ref{fig:schematic}. Research paper titles are used as queries 
in \textbf{Path 1}. The documents resulting from this search are classified 
with a paper classifier based on Random Forests~\cite{mlj01breiman}. Author names comprise the queries 
for Web search in \textbf{Path 2}, the
results of which are filtered using a 
homepage identification module trained using the RankSVM algorithm~\cite{kdd02joachims}.
The predicted academic homepages serve as seeds for 
the crawler module that obtains all documents upto a depth $2$ starting from the seed URL. 
The paper identification module is once again employed
to retain only those documents relevant to a scientific digital library among the crawled documents.
We summarize our contributions below:
\begin{itemize}
\item We propose a novel framework based on search-driven methods to 
automatically acquire research documents 
for scientific collections. To the best of our knowledge, we are the first to
use ``Web Search" to obtain seed URLs for initiating crawls 
in an open-access digital library.
%\vspace{-1mm}
\item Our Search/Crawl framework interleaves several existing and new modules.
We extend existing research on academic document classification to identify research 
papers among documents. Next, we design a novel homepage identification module,
a crucial component
 for our framework, that
uses several features based on webpage titles, URL strings, and terms in the 
result snippet to identify researcher homepages from the results of {author name search}. The identified 
homepages form seeds for our Web crawler.
%\vspace{-1mm}
\item We provide a thorough evaluation of both the paper and homepage identification components 
using various publicly-available datasets. Our proposed features 
%effectively 
attain state-of-the-art performance on both these tasks. %the paper and author homepage identification tasks.
%\vspace{-1mm}
\item Finally, we perform a large-scale, first-of-its-kind experiment using $43,496$ research paper 
titles and $32,816$ author names from Computer Science. We not only recovered approximately $75$\% of the papers
corresponding to the research paper title queries but were also able to collect about $0.665$ million 
research documents overall with our framework. These impressive yields showcase our 
Web-search driven methods to be highly effective for obtaining and maintaining up-to-date
document collections in open-access digital library portals.
\end{itemize}

%Our motivation and the Search/Crawl framework was just described in Section~\ref{sec:intro}. 
We provide details of our paper and homepage identification modules in Section~\ref{sec:methods}. In Section~\ref{sec:expts}, we describe our experimental setup, results,
and findings. We briefly summarize closely-related work in Section~\ref{sec:related} and present
concluding remarks in Section~\ref{sec:conclude}.
%Why is author search better? Because, use WWW sampling example...
%
%unless you know the crawl seeds, going to university websites and filtering is wasteful,
%WWW paper, sampled 100 webpages from 16 top universities in the US and found only 89/1600 homepages.

%\vspace{-12pt}
\section{AI Components in Our Framework}
\label{sec:methods}
The accuracy and efficiency our 
Search/Crawl framework is contingent
on the accuracies of two components: (1) the homepage identifier and (2)
the paper classifier. 

\textbf{Homepage Identification}:
Academic
homepages, known to link to research papers~\cite{nature01lawrence}, form potential seed URLs for initiating
crawls in digital libraries. For our Search/Crawl framework to be effective and efficient, it is imperative to identify 
these pages from the search results of author name queries. 
Identifying researcher homepages among other types of webpages can be treated as an instance of the 
webpage classification problem with the underlying classes: homepage/non-homepage~\cite{tweb15gollapalli}. 
However, given the Web search setting, the non-homepages retrieved in response to an author name query can be expected to be diverse with
webpages ranging from commercial websites such as LinkedIn, social media websites such as Twitter and Facebook,
publication listings such as Google Scholar, Research Gate, and several more. To handle this aspect, we draw ideas from the recent developments in 
Web search ranking and frame homepage identification as a ranking problem.

Given a set of webpages in response to a query, our objective
is to rank homepages better, i.e., top ranks, relative to other types of webpages, capturing
our preference among the webpages. For example, consider
a name query ``John Blitzer" and let 
the results in response to web search be:\\
\begin{small}
\begin{tabular}{ll}
\hline
Rank & URL \\
\hline
1 & research.google.com/pubs/author14735.html \\
2 & john.blitzer.com \\
3 & https://www.linkedin.com/pub/john-blitzer/5/606/425 \\
4 & http://dblp.uni-trier.de/pers/hd/b/Blitzer:John \\
\hline
\end{tabular}
\end{small}
\\

Suppose ``john.blitzer.com" is known to be the correct homepage and we are not interested
in other webpages. This desirable property can be expressed via
three preference pairs among the ranks:  $p_2>p_1, p_2>p_3, p_2>p_4$
where $p_i$ refers to the webpage at rank $i$. Note that, we do not express preferences
among the non-homepages $p_1$, $p_3$, and $p_4$. Preference information such as the
above is modeled through appropriate objective functions in 
learning to rank approaches~\cite{ftir09liu}. For example, a RankSVM minimizes the
Kendall’s $\tau$ measure based on the preferential ordering
information present in training examples~\cite{kdd02joachims}.

Learning to rank methods were heavily investigated for capturing user
preferences in clickthrough logs of search engines as well as in NLP tasks such as summarization and keyphrase extraction~\cite{ltrbook11li}.
Note that, unlike classification approaches that independently model both positive (homepage) and negative (non-homepage) 
classes, 
we are modeling instances in relation with each other 
with preferential ordering~\cite{kdd02joachims,icml05burges,ijcai15wan}.
We show that the ranking approach out-performs classification approaches for homepage identification
in Section~\ref{sec:expts}. We use the following feature types:
\begin{enumerate}
%\vspace{-1mm}
\item \textbf{URL Features}: Intuitively, the URL strings of academic 
homepages can be expected to contain or not contain certain
tokens. For example, a homepage URL is less likely to be hosted on domains such as ``linkedin" and ``facebook".
On the other hand, terms such as
``people" or ``home" can be expected to occur in the URL strings of homepages (example homepage URLs in Figure~\ref{fig:anecdotalsearch}) .
We tokenize the URL strings based on the ``slash (/)" separator
and the domain-name part of the URL based on the ``dot ($.$)" separator to extract 
our URL and DOMAIN feature dictionaries. 
%\vspace{-1mm}
\item \textbf{Term Features}: Current-day search engines present Web search
results as a ranked list where each webpage is indicated by its HTML title, URL string as well
as a brief summary of the content of the webpage (also known as the ``snippet"). 
Previous research has shown that
users are able to make appropriate ``click" decisions during 
Web searches based on this presented information~\cite{nips02richardson,sigir04granka}. 
We posit that users of Scholarly Web are able to identify homepages among the search results based on the term hints in titles
and snippets (for example, ``professor", ``scientist", ``student") and capture these keywords in TITLE and SNIPPET dictionaries.
%\vspace{-1mm}
\item \textbf{Name-match Features}: These features capture the common observation that 
researchers tend to use parts of their names in the URL strings of their homepages~\cite{icdm07tang,tweb15gollapalli}.
We specify two types of match features: (1) a boolean feature that indicates whether any part of the author name matches a token 
in the URL string, and
(2) a numeric feature that indicates the extent to which name tokens overlap with the (non-domain part of) URL string given by the fraction: $\frac{\#{\tt matches}}{\#{\tt name tokens}}$.
For the example author name ``Soumen Chakrabarti" and the URL string: \texttt{\small \textbf{www.cse.iitb.ac.in/$\sim$soumen}}, 
the two features have values ``true" and $0.5$, respectively.
\end{enumerate}

The dictionary sizes for the above feature types based on our training datasets (Section~\ref{sec:expts})
are listed below: \\
\begin{center}
\begin{tabular}{ll}
\hline
Feature Type & Size\\
\hline
URL+DOMAIN term features & 2025 \\
TITLE term features & 19190 \\
SNIPPET term features & 25280 \\
NAME match features & 2 \\
\hline
\end{tabular}	
\end{center}

\textbf{Paper Classification}: Recently, Caragea et al. \shortcite{iaai16caragea} studied
classification of academic documents 
into six classes: Books, Slides, Theses, Papers, CVs, and Others. 
They experimented with
bag-of-words from the textual content of the documents (BoW), tokens in the document URL string (URL),
and structural features of the document (Str) and showed that a small 
set of structural features are highly indicative of 
the class of an academic document. Their set of $43$ structural features
includes features such as size of the file, number of pages in the document, average number of words/lines
per page, phrases such as ``This thesis", ``This paper" and the relative 
position of the Introduction and Acknowledgments sections.\footnote{\scriptsize
We refer
the reader to~\cite{iaai16caragea} for a complete listing of features used
for training this classifier.}

We found that these structural features continue to perform
very well on our datasets (Section~\ref{sec:expts}) with precision/recall values in the ranges of $90+$.
Therefore, we directly employ their features for training the 
paper classification module in our framework. However, since
we are not interested in other types of documents and because binary
tasks are considered easier to learn than multiclass tasks~\cite{mlboook06bishop}, 
we re-train the classifiers for the two-class setting: papers/non-papers. 

%\vspace{-12pt}
\section{Datasets and Experiments}
\label{sec:expts}
In this section, we describe our experiments on homepage
identification and paper classification along with their performance
within the Search/Crawl 
paper acquisition framework.
Our datasets are summarized in Table~\ref{tab:dspaperauthor} and described below:
\begin{enumerate}
\item For evaluating homepage finding using author names, 
we use the
researcher homepages from DBLP, the bibliographic reference
for major Computer Science publications.\footnote{\scriptsize http://dblp.uni-trier.de/xml/}
In contrast to previous works that use this dataset to train homepage classifiers
on academic websites~\cite{tweb15gollapalli}, in
our Web search scenario, 
the non-homepages from the search results of a name query
need not be restricted to academic websites.
Except the true homepage, all other webpages 
therefore correspond to negatives. We collected the DBLP dataset as follows: 
Using the
author names as queries, we perform Web search 
and scan the 
top-$k$
results in response to each query.\footnote{\scriptsize We used the Bing API for 
all Web search experiments and retrieve the top-$10$ results. All queries are ``quoted" to 
impose exact match and ordering of tokens and the filetype syntax was used to 
retrieve PDF or HTML files as applicable.}
If the true homepage from DBLP is listed among
the top results, this URL and the others in the set of Web results can be used 
as training instances. We used RankSVM\footnote{\scriptsize http://svmlight.joachims.org/}
for learning a ranking function for author name search. In this model,
the preference among the search results for a query can be indicated by simply assigning the
ranks ``1" and ``2" respectively to the true and remaining results. 
For classification algorithms, we directly use 
the positive and negative labels for these webpages. We were able to locate homepages for $4255$ authors in the top-$10$ results 
for the author homepages listed in DBLP.
\begin{table}[htp]
\centering
\begin{small}
\begin{tabular}{|llr|}
\hline
\textbf{Dataset} & & \\ 
\hline
Research Papers &(Train) &	960(T) 472(+) \\
		&(Test)  & 959(T) 461(+) \\
\hline
DBLP Homepages &\multicolumn{2}{r|} {42,548(T) 4,255(+)} \\
\hline
CiteSeer$^x$ & \multicolumn{2}{l|} {43,496 (Titles), 32,816(Authors)} \\
\hline
\end{tabular}
\end{small}
%\vspace{-10pt}
\caption{\small Summary of datasets used in experiments. The numbers of total and positive instances are shown 
using (T) and (+), respectively, for the labeled datasets.}
\label{tab:dspaperauthor}
\end{table}
%comment-cc: we do not say what features we use for homepage classification. we should mention this somewhere. 
%\vspace{-1mm}
\item Caragea et al. \shortcite{iaai16caragea} randomly sampled
two independent sets of approximately $1000$ documents each from the crawl
data of CiteSeer$^{\tt x}$.
%focused crawler. 
%comment-cc: the detail of Train+ will only confuse the reader. It is clear that we use Train and Test sets from that paper. Also, you can keep it as "using specific search strings" but these were from the crawled documents using citeseer crawlers. Since we cite that paper, it is OK to say it here, I believe, but it is up to you.
%collected two samples of approximately 
%$1000$ academic documents each (called Train/Test)  from 
%the Web using specific search strings 
These sets, called Train and Test,  were manually labeled %with the help of two annotators 
into six classes: Paper, Book, Thesis, Slides, Resume/CV, and Others. 
%An additional dataset to handle the under-represented classes such as resumes and books in their Train dataset was also made available by them to enable a more balanced dataset for training models called ``Train+". 
We transform the documents' labels as the binary labels, Paper/Non-paper, and use these datasets directly in our experiments. %\textcolor{red}{Add more?}
%\vspace{-1mm}
\item For our third dataset,
%from the CiteSeer$^{\tt x}$ document collection. scholarly big dataset, introduced in \cite{ecir14caragea}. This scholarly big dataset is a subset of the popular open-access CiteSeer$^{\tt x}$ digital library for Computer Science and related areas and contains clean CiteSeer$^{\tt x}$ metadata records for which an entry was found in DBLP. 
%comment-cc: we extracted the 44k specifically from this set. I would cite this to emphasize that the records are more accurate than the entire citeseer collection. 
we extracted research papers from the 
publication venues listed in Table~\ref{tab:venues} from the
%{document collection of the CiteSeer$^{\tt x}$ portal.
CiteSeer$^{\tt x}$ scholarly big dataset \cite{ecir14caragea}, in which 
paper metadata (author names, venues, and paper titles) are mapped to entries in DBLP to ensure 
a clean collection.\footnote{\scriptsize Machine learning-based modules are used for extracting titles, venues, and authors of a paper 
in CiteSeer$^{\tt x}$ thus resulting in occasional erroneous metadata.} Overall, we obtained a set of $43,496$ paper titles,
authors ($32,816$ unique names) for evaluating  
our Search/Crawl framework at a large scale.

%The datasets used in training and evaluating the author and paper classifiers from web search results are listed in Table~\ref{tab:dspaperauthor}. 

\end{enumerate}

\begin{table}[!htp]
%\vspace{-10pt}
\begin{center}
\begin{scriptsize}
\begin{tabular}{|l|}
\hline	
{Total \# of research papers: 43,496, \#authors (unique names): 32,816}\\
\hline
NIPS (5211), IJCAI (4721), ICRA (3883), ICML (2979), \\ 
ACL (2970) , VLDB (2594), CVPR (2373), AAAI (2201), \\ 
CHI (2030), COLING (1933), KDD (1595), SIGIR (1454), \\
WWW (1451), CIKM (1408), SAC (1191), LREC (1128), SDM (1111), \\
EMNLP (920), ICDM (891), EACL (760), HLT-NAACL (692) \\
\hline
\end{tabular}
\end{scriptsize}
\end{center}
%\vspace{-10pt}
\caption{\small Conference venue/\#papers in the CiteSeer$^{\tt x}$ dataset. %\textcolor{red}{Can remove if required. Why don't these add up to 43496. The sum is 44522 instead}
%\vspace{-10pt}
}
\label{tab:venues}
\end{table}

We use the standard measures Precision, Recall, and F1 for summarizing the 
results of author homepage identification and paper classification~\cite{irbook08manning}. 
Unlike classification where we
consider the true and predicted labels for each instance (webpage), in RankSVM the prediction is 
per query~\cite{kdd02joachims}. That is, the results with respect to a query are assigned ranks based on scores
from the
RankSVM and the result at rank-1 is chosen as the predicted
homepage. %This results in the same value for both precision 
%and recall measures for our evaluation dataset. 
The implementations in Weka~\cite{weka}, Mallet~\cite{mallet} and SVMLight~\cite{svmlight} were used for models' training and evaluation.

%comment-cc: as a general comment in section 3, you present first the author construction set and then also first the evaluation of the author homepage identification, but in section 2, you give paper classification first. For consistency, I suggest to list them in the same order and start with author homepage. I had one of these comments from a reviewer for the IAAI paper.
\subsection{Author Homepage Finding}
We report the five-fold cross-validation performance of the homepage identification module
trained using various classification modules and RankSVM
in Table~\ref{tab:asresults}. 
The best performance obtained with all features described in Section~\ref{sec:methods}
on the DBLP dataset after tuning 
the learning parameters (such as C for SVMs), is shown in this table.
RankSVM captures the relative preferential ordering among search results and
performs the best in identifying the correct author homepage in response to a query. 
A possible reason for the lower performance of the classification approaches 
such as Binary SVMs, Na\"ive Bayes, and Maximum Entropy is 
that they model the positive and negative instances
independently and not in relation to one another for a given query. Moreover, the
diversity in webpages among the negative class is ignored and they 
are modeled uniformly as a single class in these methods.

%comment-cc: again, we need to say what features were used for the classification of homepages. 
\begin{table}[!htp]
\begin{center}
\begin{small}
\begin{tabular}{|l|c|c|c|}
\hline
\textbf{Method} & \textbf{Precision} & \textbf{Recall} & \textbf{F1} \\
\hline
Na\"ive Bayes & 0.4830 &   0.9239 &   0.63432 \\
MaxEnt &  0.8207 &   0.8002 &   0.8102 \\
Binary SVM & 0.8353 &   0.8149 &   0.8249 \\ %c=0.1
RankSVM & \textbf{0.8900} & \textbf{0.8900} & \textbf{0.8900} \\ %c=0.01.out
%MaxEnt &  0.8151 &  0.7699 &   0.7918 \\
%Na\"ive Bayes & 0.5477 &   0.8654 &   0.6707 \\
%Binary SVM &  0.7557 &   0.7975 &  0.7759 \\
%RankSVM & \textbf{0.8924} & \textbf{0.8924} & \textbf{0.8924} \\
\hline
\end{tabular}
\end{small}
\end{center}
%\vspace{-10pt}
\caption{\small Classifier and RankSVM performances on DBLP dataset.}
\label{tab:asresults}
\end{table}

We point out that false positives are not very critical in our Search/Crawl framework.
Including an incorrectly predicted homepage as
a seed URL may result in crawling irrelevant documents and extra processing load. However, 
these documents are subsequently filtered out by our paper classifier.

\begin{table}[!htp]
\centering
\begin{small}
\begin{tabular}{l r || l r}
\hline
\textbf{FeatureType} & \textbf{Feature} & \textbf{FeatureType} & \textbf{Feature} \\ \hline
NAME & fracMatch & TITLE & university \\  
DOMAIN & com & SNIPPET & computer \\ 
NAME & hasMatch & TITLE & homepage \\  
TITLE & home & SNIPPET & university \\ 
TITLE & page & TITLE & linkedin \\  
SNIPPET & professor & SNIPPET & science \\  
DOMAIN & edu & SNIPPET & discover \\  
SNIPPET & view & URL & author \\  
SNIPPET & department & SNIPPET & linkedin \\ 
SNIPPET & profile & SNIPPET & professionals \\ \hline 
\end{tabular}
\end{small}
%\vspace{-10pt}
\caption{\small The top-$20$ features ranked based on Information Gain.}
\label{tab:asigfeats}
\end{table}
Table~\ref{tab:asigfeats} shows the top features based on information gain values~\cite{jmlr03forman03}.
These features make intuitive sense; for instance, a researcher homepage is likely to have parts of 
the researcher name mentioned on it along with terms like ``home" and ``page" in the HTML title. 
Similarly, webpages typically ending in ``.com" or 
having ``linkedin" in their description are unlikely to be homepages.
	
\subsection{Research Paper Identification}
The results of paper classification are summarized in Table~\ref{tab:perfpaper}.
We directly used the feature sets proposed by Caragea et al. \shortcite{iaai16caragea}
and tested various classifiers including Na\"ive Bayes, Support Vector Machines and 
Random Forests. All models are trained on the ``Train'' dataset. The parameters of each model are tuned 
through cross-validation on the ``Train" dataset and the classification performance evaluated
on the ``Test" dataset. {The results of various
features sets using a Random Forest for the ``paper" class in the binary setting
are shown in Table~\ref{tab:perfpaper}. We also show
the performance on the ``paper" class with the 
multiclass setting and the weighted averages of all measures over all classes for
both the settings in this table. 
The best classification performance is obtained using a Random Forest trained on
structural features with the overall performance being
substantially better in the two-class setting rather than the multiclass setting.}
\begin{table}[htp]
\centering
\begin{small}
%\begin{tabular}{|l|c|c|c|c|}
%\hline
%{Feature}&{Precision}&{Recall}&{F1}&{Accuracy}\\
%\hline
%BoW (P)&  0.86  &    0.92 &  0.889 & 88.94\%  \\
%URL (P)& 0.729 & 0.729 & 0.729 & 73.93\%\\
%Str/Binary (P)& {\bf 0.933}  &    {\bf 0.967} &  {\bf 0.950}  & {\bf 95.09\%}  \\
%\hline
%Str/Multiclass (P) & 0.918  &  0.965   &  0.941  & 90.19\%  \\
%\textcolor{red}{Str/Binary (A)}& -  &  -   & -   &  - \\
%\textcolor{red}{Str/Multiclass (A)}& -  &  -   & -   & -  \\

\begin{tabular}{|l|c|c|c|}
\hline
{Feature}&{Precision}&{Recall}&{F1}\\
\hline
BoW (P)&  0.86  &    0.92 &  0.889   \\
URL (P)& 0.729 & 0.729 & 0.729 \\
Str/Binary (P)& \textbf{0.933}  &  \textbf{0.967}  &  \textbf{0.950} \\
Str/Multiclass (P) & 0.918 & 0.965 &  0.941   \\
\hline
Str/Binary (A)& \textbf{0.952}  &   \textbf{0.951}  & \textbf{0.951}  \\
Str/Multiclass (A) & 0.893 &     0.902 &  0.892  \\
\hline
\end{tabular}
\end{small}
%\vspace{-10pt}
\caption{\small Classification performance on the test dataset. `P/A' indicate performances for ``Paper"/``All" classes.}
\label{tab:perfpaper}
\end{table}

\subsection{Search/Crawl Experiments}
\begin{table*}[!htp]
\begin{center}
\begin{tabular}{|l|c|c|c|c|}
\hline
\#Queries & \#PDFs & \#Papers & \#UniqueTitles & \#Matches\\
\hline
43,496 titles (Path 1)& 322,029 & 213,683 & 91,237 & 32,565 \\
%%%EMAIL on 1/2/2016 from Krutarth, In Table 6, #Matches in Path 1 ,
%%%We have total 32,768 matching without removing anything.
%%%We have total 32,702 matching after removing files for which we don't have URL mapping.
%%%We have total 32,565 matching after removing files for which we don't have URL mapping, and also removing files from CiteSeer.
32,816 names (Path 2) & 665,661 & 452,273 & 204,014 & 17,627\\
\hline
\end{tabular}
%\vspace{-10pt}
\caption{\small \#Papers obtained through the two paths in our Search/Crawl framework.}
%\vspace{-12pt}
\label{tab:csxresults}
\end{center}
\end{table*}
Finally, we evaluate the two AI components in 
practice within our Search/Crawl framework
using the CiteSeer$^{\tt x}$ dataset. To this end, for \textbf{Path 1}, we use
the $43,496$ paper titles as search queries. Structural features
extracted 
from the resulting PDF documents of each search are used
to identify research documents with our paper classifier. For \textbf{Path 2}, the $32,816$ author names are used as queries. The
RankSVM-predicted homepages from the results of each query are crawled for  PDF documents up to a depth of 
$2$.\footnote{\scriptsize We used the wget utility for our crawls (https://www.gnu.org/software/wget/)} 
Once again, the paper classifier is employed to
identify research documents among the crawled documents. 

The number of PDFs and papers found through the two paths in 
our proposed Search/Crawl framework are shown in Table~\ref{tab:csxresults}. 
Since our dataset is based on CiteSeer$^{\tt x}$, we removed all paper search results
that point to CiteSeer$^{\tt x}$ URLs for a fair evaluation. The number of papers 
that we could obtain from the original $43,496$ collection through both the paths 
are shown in the last column of this table. We use the title
and author names available in the dataset to look up the first page of the PDF document
for computing this match.

We are able to obtain $75\%$ ($\frac{32565}{43496}$)
of the original titles through \textbf{Path 1} compared to the $40\%$ 
($\frac{17627}{43496}$) through \textbf{Path 2} (column $5$ in Table~\ref{tab:csxresults}).
%Based on these numbers, it appears that a targeted Web search using specific title keywords of the paper
%may be more appropriate
%for finding specific papers. 
{In general,
given that paper titles contain representative keywords~\cite{emnlp14caragea,Litvak2008},
if they are available online, a Web
search with appropriate filetype filters
is a successful strategy for finding them. The high percentage
of papers found along  \textbf{Path 2} confirms previous findings that
researchers tend to link to their papers via their homepages~\cite{nature01lawrence,tweb15gollapalli}.}

Intuitively, the overall yield can be expected to be higher through \textbf{Path 2}.
Once an author homepage is reached, other research papers linked to this page 
can be directly obtained. Indeed, as shown in columns $2$ and $3$ of Table~\ref{tab:csxresults}, 
the numbers of PDFs as well as classified papers are significantly higher along \textbf{Path 2}. 
Crawling the predicted homepages of the $32,816$ authors we obtain approximately
$14$ research papers per query on average ($\frac{452273}{32816}=13.78$). In contrast,
examining only the top-$10$ search results along \textbf{Path 1}, we obtain $5$ research 
documents per query ($\frac{213683}{43496}=4.91$). 

We used the CRF-based title extraction tool for research papers, ParsCit~\footnote{\scriptsize http://aye.comp.nus.edu.sg/parsCit/} 
to extract the titles of the research papers obtained from both the paths. 
%Though this tool 
%is not very accurate [TODO add ECIR citation, if possible mention number], 
%we use it to estimate the number of unique titles obtained through each path of our framework.
{The number of extracted unique titles are shown in column $4$ of Table~\ref{tab:csxresults}. 
The overlap in the two sets of titles is $28,374$. Compared to the overall yields
along \textbf{Path 1} and \textbf{Path 2},
this small overlap indicates that the two paths are capable of reaching different sections of the Web and play
complementary roles in our framework. For example, the top-$20$ domains of the URLs from which 
we obtained research papers along \textbf{Path 1} are shown in Table~\ref{tab:topdomainsp1}. Indeed, 
via Web search we are able to reach a wide range of domains. This is unlikely in crawl-driven methods
without an exhaustive list of seeds since only
links up to a specified depth from a given seed are explored~\cite{irbook08manning}.}

To summarize, using about $0.076$ million queries ($43,496+32,816$) in our framework, we are able to build a
collection of $0.665$ million research documents ($213,683+452,273$) and 
$0.267$ million unique titles
($91,237+204,014-28,374$). About $32-33\%$ of the obtained documents
are ``non-papers" along both the paths. Scholarly Web is known to contain a variety of documents
including project proposals, resumes, and course materials~\cite{jis06ortega}. 
Indeed, some of these documents may include the exact paper titles and show up in paper search results as well
as be linked to author homepages. In addition, 
using incorrectly-predicted homepage as seeds may result in ``bad" documents.
\begin{table}[!hp]
\centering
\begin{small}
\begin{tabular}{|l|}
\hline
edu (71,139), org (47,272), net (20,552), com (19,178), de (5,424)\\
uk  (5,065), fr (3,770), ca (3,651), it (2,647), gov (2,130), \\
nl  (1,891), cn (1,777), jp (1,673), au (1,655), cc (1,489), \\ 
ch (1,431), sg  (1,282), in (1,209), il (1,206), es (1,144) \\
\hline
\end{tabular}
\end{small}
%\vspace{-10pt}
\caption{\small The top-20 domains from which papers were obtained along Path-1 of our framework.}
\label{tab:topdomainsp1}
\end{table}
%comment-cc: I would keep this figure or a table to show where the yield comes from.

\textbf{Sample Evaluation.} Given the size of the CiteSeer$^{\tt x}$ dataset and the 
large number of documents obtained via the Search/Crawl framework (Table~\ref{tab:csxresults}), it is
extremely labor-intensive to manually examine all documents resulting from this experiment. However,
since our classifiers and rankers are not $100\%$ accurate and we only
examine the top-$k$ results from the search engine, we need an estimate of how many papers
we are able to obtain via our Search/Crawl approach among those
that are actually obtainable on the Web. We randomly selected $10$ titles from the CiteSeer$^{\tt x}$ dataset and their 
associated set of $78$ authors and inspected 
all PDFs that can be obtained via our search/crawl framework manually.
That is, we searched for the selected paper titles and manually examineed and annotated the resulting PDFs.
Similarly, the correct homepages of the $78$ authors were obtained by searching the Web
and manually examining the resulting webpages. The correct homepages were crawled (to depth 2)
for PDFs and the resulting documents were manually labeled.

We were able to locate $49$ correct homepages of the $78$ authors in this manual experiment. Crawling these homepages resulted
in $2116$ PDFs out of which $1418$ were found to be research papers.
Our Search/Crawl framework that crawls predicted homepages for the $78$ authors 
and uses paper classifier
predictions to identify research papers was able to 
obtain $1291$ research papers. Out of these documents, $1104$ match with
the intended set of $1418$ papers. Thus, we are able to 
obtain approximately $\textbf{78}\%$ of the intended set of papers along with an additional $187$ new ones.
Paper search using titles results in $59$ PDFs out of which $33$ are true papers.
Our paper classifier obtains a precision/recall of $\textbf{84}$\%/$\textbf{97}$\%, predicting $32$ out of these $33$ papers correctly
and $38$ papers overall. 

%\vspace{-12pt}
\section{Related Work}
\label{sec:related}
Homepage finding and document classification are 
very well-studied problems. 
Due to space constraints, we refer the reader to 
the TREC 2001 proceedings\footnote{\scriptsize http://trec.nist.gov/proceedings/proceedings.html}
and the comprehensive
reviews of the feature representations, methods, and results
for various text/webpage classification problems~\cite{cs02sebastiani,cs09qi}.

Though homepage finding in TREC did not
specifically address researcher homepages, this track 
resulted in various state-of-the-art machine learning systems
for finding homepages~\cite{spire02xi,tis03upstill,icadl06wang}. 
%. Indeed, several query dependent and independent features such as 
%query-content match, PageRank, anchor text of incoming links etc. were 
%previously explored for this task
Among works focusing specifically on researcher homepages, both Tang et al.~\shortcite{icdm07tang} and Gollapalli et al.~\shortcite{tweb15gollapalli}
treat homepage finding as a binary classification task and use
various URL and content features. Ranking methods were 
explored for homepage finding using the top terms
obtained from topic models~\cite{sigireos11gollapalli}.

In the context of scientific digital libraries,
document classification into
classes related to subject-topics (for example, ``machine learning", ``databases")
was studied previously~\cite{icml03lu,emnlp15caragea}.
Often bag-of-words features as well as topics extracted using LDA/pLSA are used to represent
the underlying documents in these works. Structural features, on the other
hand, are popular in classifying and clustering semi-structured XML documents~\cite{icpr08ghosh,aireview13asghari}. 
%\textcolor{red}{[TODO should we cite WSC work here?]}
%comment-cc: I would not cite this paper, at least not for this version as may cause confusion and we already have a good discussion / comparison with the 6-class classification.

In contrast with existing work, %we show that structural features are more
%reliable for identifying
%research papers. In addition, we are the first to 
we investigate features from web search engine results %(such as URL strings and snippets) 
and formulate researcher homepage identification as a learning to rank task. In addition,
we are the first to interleave various AI components with 
existing Web search and crawl modules to build an efficient paper acquisition framework.

%\vspace{-12pt}
\section{Conclusions}
\label{sec:conclude}
We proposed a search-driven framework for automatically acquiring research
documents on the Web as an alternative to
crawl-driven methods adopted in current open-access digital libraries.
Our framework crucially depends on accurate paper classification and researcher 
homepage identification modules. To this end, we discussed features for 
these modules and showed experiments illustrating their
state-of-the-art performance. In one experiment using a large collection
of about $0.076$ million queries, our framework was able to automatically
acquire a collection of approximately $0.665$ million research documents. These
results showcase the potential of our proposed framework in 
improving scientific digital library collections.
{For future work, apart from improving the accuracies of individual components in our framework, we will focus on 
including other document formats (for example, .ps and zipped files)
as well as other document types (for example, course materials).}

\section*{Acknowledgments}

We are grateful to Dr. C. Lee Giles for the CiteSeerX data. We also thank Corina Florescu and Kishore Neppalli for their help with various dataset construction tasks. This research was supported in part by the NSF award \#1423337 to Cornelia Caragea. Any opinions, findings, and conclusions expressed here are those of the authors and do not necessarily reflect the views of NSF.

%\vspace{-10pt}
\small
\bibliographystyle{named}

\begin{thebibliography}{}

\bibitem[\protect\citeauthoryear{Asghari and
  KeyvanPour}{2013}]{aireview13asghari}
Elaheh Asghari and MohammadReza KeyvanPour.
\newblock Xml document clustering: techniques and challenges.
\newblock {\em Artificial Intelligence Review}, 2013.

\bibitem[\protect\citeauthoryear{Balog and De~Rijke}{2007}]{ijcai07balog}
Krisztian Balog and Maarten De~Rijke.
\newblock Determining expert profiles (with an application to expert finding).
\newblock In {\em IJCAI}, 2007.

\bibitem[\protect\citeauthoryear{Bishop}{2006}]{mlboook06bishop}
Christopher~M. Bishop.
\newblock {\em Pattern Recognition and Machine Learning (Information Science
  and Statistics)}.
\newblock Springer-Verlag New York, Inc., 2006.

\bibitem[\protect\citeauthoryear{Breiman}{2001}]{mlj01breiman}
Leo Breiman.
\newblock Random forests.
\newblock In {\em Machine Learning}, volume~45, pages 5--32, 2001.

\bibitem[\protect\citeauthoryear{Broder}{2002}]{sigir02broder}
Andrei Broder.
\newblock A taxonomy of web search.
\newblock {\em SIGIR Forum}, 36(2), September 2002.

\bibitem[\protect\citeauthoryear{Burges \bgroup \em et al.\egroup
  }{2005}]{icml05burges}
Chris Burges, Tal Shaked, Erin Renshaw, Ari Lazier, Matt Deeds, Nicole
  Hamilton, and Greg Hullender.
\newblock Learning to rank using gradient descent.
\newblock In {\em ICML}, 2005.

\bibitem[\protect\citeauthoryear{Caragea \bgroup \em et al.\egroup
  }{2015}]{emnlp15caragea}
Cornelia Caragea, Florin Bulgarov, and Rada Mihalcea.
\newblock Co-training for topic classification of scholarly data.
\newblock In {\em EMNLP}, 2015.

\bibitem[\protect\citeauthoryear{Caragea \bgroup \em et al.\egroup
  }{2016}]{iaai16caragea}
Cornelia Caragea, Jian Wu, Sujatha~Das Gollapalli, and C.~Lee Giles.
\newblock Document type classification in online digital libraries.
\newblock {\em Innovative Applications of Artificial Intelligence (IAAI)},
  2016.

\bibitem[\protect\citeauthoryear{Forman}{2003}]{jmlr03forman03}
George Forman.
\newblock An extensive empirical study of feature selection metrics for text
  classification.
\newblock {\em J. Mach. Learn. Res.}, 3:1289--1305, 2003.

\bibitem[\protect\citeauthoryear{Ghosh and Mitra}{2008}]{icpr08ghosh}
S.~Ghosh and P.~Mitra.
\newblock Combining content and structure similarity for xml document
  classification using composite svm kernels.
\newblock In {\em ICPR}, 2008.

\bibitem[\protect\citeauthoryear{Gollapalli \bgroup \em et al.\egroup
  }{2011}]{sigireos11gollapalli}
Sujatha~Das Gollapalli, Prasenjit Mitra, and C.~Lee Giles.
\newblock Learning to rank homepages for researcher name queries.
\newblock In {\em EOS Workshop at SIGIR}, 2011.

\bibitem[\protect\citeauthoryear{Gollapalli \bgroup \em et al.\egroup
  }{2015}]{tweb15gollapalli}
Sujatha~Das Gollapalli, Cornelia Caragea, Prasenjit Mitra, and C.~Lee Giles.
\newblock Using unlabeled data to improve researcher homepage classification.
\newblock In {\em Transactions on the Web}, 2015.

\bibitem[\protect\citeauthoryear{Granka \bgroup \em et al.\egroup
  }{2004}]{sigir04granka}
Laura~A. Granka, Thorsten Joachims, and Geri Gay.
\newblock Eye-tracking analysis of user behavior in www search.
\newblock In {\em SIGIR}, 2004.

\bibitem[\protect\citeauthoryear{Hall \bgroup \em et al.\egroup }{2009}]{weka}
Mark Hall, Eibe Frank, Geoffrey Holmes, Bernhard Pfahringer, Peter Reutemann,
  and Ian~H. Witten.
\newblock The weka data mining software: An update.
\newblock In {\em SIGKDD Explorations}, 2009.

\bibitem[\protect\citeauthoryear{He \bgroup \em et al.\egroup
  }{2009}]{cikm09he}
Qi~He, Bi~Chen, Jian Pei, Baojun Qiu, Prasenjit Mitra, and C.~Lee Giles.
\newblock Detecting topic evolution in scientific literature: how can citations
  help?
\newblock {\em CIKM}, 2009.

\bibitem[\protect\citeauthoryear{He \bgroup \em et al.\egroup
  }{2011}]{wsdm11he}
Qi~He, Daniel Kifer, Jian Pei, Prasenjit Mitra, and C.~Lee Giles.
\newblock Citation recommendation without author supervision.
\newblock In {\em WSDM}, 2011.

\bibitem[\protect\citeauthoryear{Joachims}{1999}]{svmlight}
T.~Joachims.
\newblock Making large-scale {SVM} learning practical.
\newblock In B.~Sch{\"o}lkopf, C.~Burges, and A.~Smola, editors, {\em Advances
  in Kernel Methods - Support Vector Learning}, chapter~11, pages 169--184. MIT
  Press, Cambridge, MA, 1999.

\bibitem[\protect\citeauthoryear{Joachims}{2002}]{kdd02joachims}
Thorsten Joachims.
\newblock Optimizing search engines using clickthrough data.
\newblock In {\em SIGKDD}, 2002.

\bibitem[\protect\citeauthoryear{Kataria \bgroup \em et al.\egroup
  }{2011}]{ijcai11kataria}
Saurabh Kataria, Prasenjit Mitra, Cornelia Caragea, and C.~Lee Giles.
\newblock Context sensitive topic models for author influence in document
  networks.
\newblock In {\em IJCAI}, 2011.

\bibitem[\protect\citeauthoryear{Lawrence}{2001}]{nature01lawrence}
Steve Lawrence.
\newblock Free online availability substantially increases a paper's impact.
\newblock In {\em Nature}, 411 (6837), pages 521--521, 2001.

\bibitem[\protect\citeauthoryear{Li \bgroup \em et al.\egroup
  }{2006}]{infoscale06li}
Huajing Li, Isaac~G. Councill, Levent Bolelli, Ding Zhou, Yang Song, Wang-Chien
  Lee, Anand Sivasubramaniam, and C.~Lee Giles.
\newblock Citeseer$^{\tt x}$: a scalable autonomous scientific digital library.
\newblock {\em InfoScale}, 2006.

\bibitem[\protect\citeauthoryear{Li}{2011}]{ltrbook11li}
Hang Li.
\newblock {\em Learning to Rank for Information Retrieval and Natural Language
  Processing}.
\newblock Morgan \& Claypool Publishers, 2011.

\bibitem[\protect\citeauthoryear{Liu}{2009}]{ftir09liu}
Tie-Yan Liu.
\newblock Learning to rank for information retrieval.
\newblock {\em Found. Trends Inf. Retr.}, 2009.

\bibitem[\protect\citeauthoryear{Lu and Getoor}{2003}]{icml03lu}
Ching Lu and Lise Getoor.
\newblock Link-based classification.
\newblock In {\em ICML}, 2003.

\bibitem[\protect\citeauthoryear{Manning \bgroup \em et al.\egroup
  }{2008}]{irbook08manning}
Christopher~D. Manning, Prabhakar Raghavan, and Hinrich Schutze.
\newblock {\em Introduction to Information Retrieval.}
\newblock Cambridge University Press, New York, NY, USA., 2008.

\bibitem[\protect\citeauthoryear{McCallum}{2002}]{mallet}
Andrew~Kachites McCallum.
\newblock Mallet: A machine learning for language toolkit.
\newblock http://mallet.cs.umass.edu, 2002.

\bibitem[\protect\citeauthoryear{Ortega-Priego \bgroup \em et al.\egroup
  }{2006}]{jis06ortega}
José-Luis Ortega-Priego, Isidro~F. Aguillo, and José~Antonio Prieto-Valverde.
\newblock Longitudinal study of contents and elements in the scientific web
  environment.
\newblock {\em Journal of Information Science}, 32(4), 2006.

\bibitem[\protect\citeauthoryear{Qi and Davison}{2009}]{cs09qi}
Xiaoguang Qi and Brian~D. Davison.
\newblock Web page classification: Features and algorithms.
\newblock {\em ACM Comput. Surv.}, 41(2), February 2009.

\bibitem[\protect\citeauthoryear{Richardson and
  Domingos}{2002}]{nips02richardson}
Mathew Richardson and Pedro Domingos.
\newblock The {I}ntelligent {S}urfer: {P}robabilistic {C}ombination of {L}ink
  and {C}ontent {I}nformation in {P}age{R}ank.
\newblock In {\em NIPS}, 2002.

\bibitem[\protect\citeauthoryear{Sebastiani}{2002}]{cs02sebastiani}
Fabrizio Sebastiani.
\newblock Machine learning in automated text categorization.
\newblock {\em ACM Comput. Surv.}, 34(1), 2002.

\bibitem[\protect\citeauthoryear{Serdyukov \bgroup \em et al.\egroup
  }{2008}]{cikm08serdyukov}
Pavel Serdyukov, Henning Rode, and Djoerd Hiemstra.
\newblock Modeling multi-step relevance propagation for expert finding.
\newblock In {\em CIKM}, 2008.

\bibitem[\protect\citeauthoryear{Tang \bgroup \em et al.\egroup
  }{2007}]{icdm07tang}
Jie Tang, Duo Zhang, and Limin Yao.
\newblock Social network extraction of academic researchers.
\newblock In {\em ICDM}, 2007.

\bibitem[\protect\citeauthoryear{Tang \bgroup \em et al.\egroup
  }{2008}]{kdd08tang}
Jie Tang, Jing Zhang, Limin Yao, Juanzi Li, Li~Zhang, and Zhong Su.
\newblock Arnetminer: extraction and mining of academic social networks.
\newblock {\em KDD}, 2008.

\bibitem[\protect\citeauthoryear{Upstill \bgroup \em et al.\egroup
  }{2003}]{tis03upstill}
Trystan Upstill, Nick Craswell, and David Hawking.
\newblock Query-independent evidence in home page finding.
\newblock {\em ACM Trans. Inf. Syst.}, 2003.

\bibitem[\protect\citeauthoryear{Wan and Xiao}{2008}]{aaai08wan}
Xiaojun Wan and Jianguo Xiao.
\newblock Single document keyphrase extraction using neighborhood knowledge.
\newblock In {\em AAAI}, 2008.

\bibitem[\protect\citeauthoryear{Wan \bgroup \em et al.\egroup
  }{2015}]{ijcai15wan}
Ji~Wan, Pengcheng Wu, Steven C.~H. Hoi, Peilin Zhao, Xingyu Gao, Dayong Wang,
  Yongdong Zhang, and Jintao Li.
\newblock Online learning to rank for content-based image retrieval.
\newblock In {\em IJCAI}, 2015.

\bibitem[\protect\citeauthoryear{Wang and McCallum}{2006}]{kdd06wang}
Xuerui Wang and Andrew McCallum.
\newblock Topics over time: A non-markov continuous-time model of topical
  trends.
\newblock In {\em KDD}, 2006.

\bibitem[\protect\citeauthoryear{Wang and Oyama}{2006}]{icadl06wang}
Yuxin Wang and Keizo Oyama.
\newblock Web page classification exploiting contents of surrounding pages for
  building a high-quality homepage collection.
\newblock In {\em ICADL}, 2006.

\bibitem[\protect\citeauthoryear{Xi \bgroup \em et al.\egroup
  }{2002}]{spire02xi}
Wensi Xi, Edward~A. Fox, Roy~P. Tan, and Jiang Shu.
\newblock Machine learning approach for homepage finding task.
\newblock In {\em String Processing and Information Retrieval}, 2002.

\end{thebibliography}

\end{document}